\documentstyle[11pt,newpasp,twoside,psfig ]{article}
\index{summary}
\index{instructions}
\index{template}

\def\edcomment#1{\iffalse\marginpar{\raggedright\sl#1\/}\else\relax\fi}
\reversemarginpar

\begin{document}
\title{Rejuvenating the Shells of Supernova Remnants by Pulsar Winds}
 \author{Eric van der Swaluw}
\affil{Dublin Institute for Advanced Studies, 5 Merrion Square, Dublin 2,
Ireland}
\author{Abraham Achterberg}
\affil{Astronomical Institute, Utrecht University, P.O.Box 80000, 3508 TA
Utrecht, The Netherlands}
\author{Yves A. Gallant}
\affil{Service d'Astrophysique, CEA Saclay, 91191 Gif-sur-Yvette Cedex, France}

\begin{abstract}
We reconsider the rejuvenation mechanism as proposed by Shull, Fesen, \& Saken
(1989). These authors suggest that an active pulsar can catch up with, and
rejuvenate the shell of the associated supernova remnant. The morphology
of the SNRs G5.4-1.2 and CTB80 seem to confirm this rejuvenation mechanism. The
spindown energy is deposited by the pulsar as a relativistic pulsar wind, and 
has a sufficient power to explain the observed radio emission
observed in these remnants. Shull et al. (1989) did {\it not} explain the observed
lengthscales of the rejuvenated parts of the SNR shell. therefore one needs to
consider the diffusive transport of the injected electrons by the pulsar wind.
We propose to apply a diffusion mechanism as introduced by Jokipii (1987), which
makes a distinction between diffusion along the magnetic field lines and
perpendicular to the magnetic field lines, parameterised by the gyro factor
$\eta$. We show that one has to assume a high value for the gyro factor,
$\eta\simeq 10^3-10^4$, i.e. diffusion of the electrons along the magnetic field
line is much faster then perpendicular to the magnetic field line, in order for
the rejuvenation mechanism to work on the observed lengthscales.
\end{abstract}

\section{Introduction}

A supernova remnant (SNR) results from the supernova explosion of a massive
star. When the core of the progenitor star collapses, a neutron star can be
formed. Several mechanisms proposed for a core collapse supernova can impart
a kick velocity to the pulsar, although no particular mechanism can be 
favoured.

The relativistic pulsar wind will interact with the SNR, resulting in a pulsar
wind nebula (PWN), which contains the shocked pulsar wind material. Initially
the PWN will be centrally located in the SNR, but due to the kick velocity of
the pulsar, the PWN will be dragged along by the pulsar, being deformed into a 
bow shock and ultimately break through the shell of the decelerating SNR 
(van der Swaluw, Achterberg, \& Gallant 1998; Chevalier 1998). 

Figure 1 depicts the configuration of this last interacting stage of such a
composite remnant. The relativistic pulsar wind is terminated by a strong
MHD termination shock, whereas the bubble around 
the termination shock has been 
deformed into a bow  shock due to the supersonic motion of the pulsar. 
Relativistic particles are injected at the site of the termination shock, of 
which a large fraction is advected away resulting in a wake of relativistic
particles, which explains the observed trail of radio emission for bow shock
PWN. However, due to the diffusive transport of the injected particles,
a small fraction of these particles will radiate part of their energy away in 
the magnetic field of the SNR shell. Shull et al. (1989) have argued that these
freshly injected particles brighten the radio emission of the SNR shell.
 
We calculate the diffusive length scales of the injected electrons in the
lifetime of a SNR or their radiative lifetime. This will enable us to compare
these results with actual observed lengthscales in these systems. We will show
that we need strongly anisotropic diffusion in order to obtain agreement
between our calculations and the observed properties of several composite
remnants.

\begin{figure*}[ht]
\begin{center}
\centerline{\psfig{file=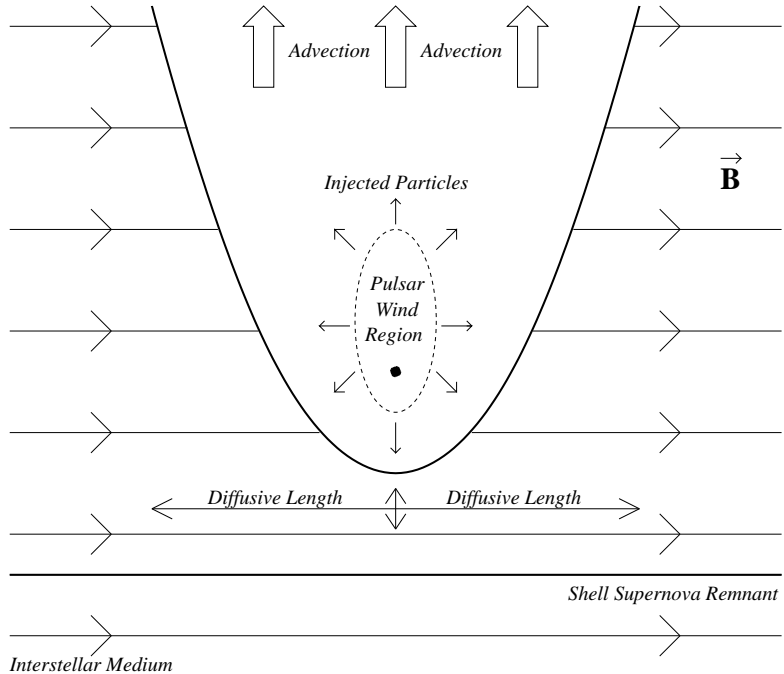,height=3.5in}}
\end{center}
\caption{Configuration of a pulsar wind interacting with the shell of a
supernova remnant. The pulsar wind is terminated by a strong MHD shock (dashed
line) and the PWN itself is bounded by a bow shock (solid line). The PWN is
propagating through the shell of its associated SNR, where the magnetic field
lines are parallel with the shell of the SNR. By considering anisotropic
diffusion, one cane explain the lengthscales of rejuvenated shells of SNRs 
like G5.4-1.2 and CTB80.}
\label{fig:1}
\end{figure*}

\section{Synchrotron theory and diffusive transport}

\subsection{General theory}

Relativistic electrons in an astrophysical flow radiate part of their energy
away as synchrotron radiation, due to the presence of a magnetic field, $B$ in
the plasma. Using standard synchrotron theory (see e.g. Rybicky \& Lightman
1979), one can write the frequency where the emission peaks as:
\begin{equation}
        \nu_{\rm MHz} \; \simeq \; 4.67 B_{\mu G} E_{\rm GeV}^2\;\;\;{\rm MHz},
\end{equation}
here $E_{\rm GeV}$ is the energy of the electron in GeV. The timescale on 
which the electron has lost half of its energy due to synchrotron losses 
can be written as:
\begin{equation}
\label{Synchr}
        \tau_{\rm loss} \; \simeq \; {2\times 10^9 B_{\mu {\rm G}}^{-3/2} 
                          \nu_{\rm MHz}^{-1/2} \;\;\; {\rm yr}}.
\end{equation}
The propagation of relativistic particles through the flow of a plasma is a
combination of advection by the large-scale flow, and diffusion with respect 
to this flow. We will consider the limit of Bohm diffusion, in which case the
mean free path $\lambda$ equals the gyroradius $r_{\rm g}$ of the particle.
Using this limit, one can rewrite the Bohm diffusion coefficient $\kappa_{\rm
B}$ in terms of
the above characteristic frequency $\nu_{\rm MHz}$, which yields:
\begin{equation}
        \kappa_{\rm B} \simeq 7.5\times 10^{21} \nu_{\rm MHz}^{1/2}
                               B_{\mu{\rm G}}^{-3/2} \;\;\; {\rm cm^2/sec}.
\end{equation}
Now by considering particles injected at $t=0$, one can write the diffusion
lengthscale $\Delta R_{\rm syn}$, at the synchrotron loss time 
$\tau_{\rm loss}$ in the case for Bohm diffusion as:
\begin{equation}
\label{DiffL}
        \Delta R_{\rm syn} =\sqrt{2\kappa_{\rm B} \tau_{\rm loss}}\simeq
         9.7 B_{\mu{\rm G}}^{-3/2} \;\;\; {\rm parsec}.
\end{equation}
The above
equation will serve as a reference value, when we consider the more general
case of diffusion in the next subsection, where the mean free path $\lambda$ 
satisfies $\lambda\gg r_{\rm g}$. 

\subsection{Application to pulsar winds}

We consider the case of a pulsar wind close to the shell of a SNR. Because of
the small size of a PWN bow shock ($\sim 0.1$ parsec) compared with the SNR
($\sim 10.0$ parsec), we can approximate the site where the relativistic
electrons are injected as a point source in the SNR interior. We can calculate
the diffusion lengthscale again, this time using the crossing time for a pulsar
with speed $V_{\rm psr}$ to catch up with the SNR shell with radius $R_{\rm
snr}$, i.e. $R_{\rm snr}/V_{\rm psr}$. Using Bohm diffusion again yields:
\begin{equation}
	\Delta R_{\rm radio} \simeq 2.2\times 10^{-2} \nu_{\rm MHz}^{1/4}
			       B_{\mu{\rm G}}^{-3/4} 
			      ( R_{\rm snr}/V_{\rm psr})^{1/2}
			       \;\;\; {\rm parsec}.
\end{equation}
From the above equation one can see, that using standard Bohm diffusion, {\it
the obtained diffusive lengthscale is similar to the size of the PWN itself}, 
i.e. the radio electrons are confined to the surroundings of the PWN. However, 
the observed size of the rejuvenated shells is much larger ($\sim$ 10 parsec). 
therefore we have to extend the diffusion model in order for the diffusive 
lengthscales to match the observed lengthscales.

We follow Jokipii (1987) by making a distinction between diffusion along the
magnetic field lines ($\kappa_\parallel$) and perpendicular to the magnetic 
field lines ($\kappa_\perp)$. 
The mean free path along the magnetic field line then equals
$\lambda_{\parallel}=\eta r_{\rm g}$, with $\eta$ the gyro factor, which is 
related to the turbulence level $\delta B$ in the magnetic field by 
$\eta =(\delta B/B)^{-2}$.
The diffusion coefficient along the magnetic field lines equals $
\kappa_{\parallel} = \eta \kappa_{\rm B}$. One usually assumes $\eta\gg 1$.
Perpendicular to the magnetic field lines we follow Jokipii (1987) and assume
that a particle scatters one gyroradius across field lines for every parallel
scattering length, so $\kappa_{\perp} = \eta \kappa_{\rm B}/(1+\eta^2)
	\simeq\kappa_{\rm B}/\eta$.
With this description for diffusion, particles diffuse faster along the magnetic
field lines then compared with quasi-isotropic Bohm 
diffusion.
Using this description together with the timescale for synchrotron losses, 
$\tau_{\rm loss}$, one can write the diffusion lengthscale at time $t$ of 
the radio electrons as:
\begin{equation}
\label{Diffusion}
	\Delta R_{\parallel} = \Delta R_{\rm syn}\times 
	{\left({\kappa_{\parallel}}\over\kappa_{\rm B}\right)}^{1/2}
	\times\left({t\over\tau_{\rm loss}}\right)^{1/2},
\end{equation}
with $\tau_{\rm loss}$ given by equation (\ref{Synchr}) and 
$\Delta R_{\rm syn}$ given by equation (\ref{DiffL}). The above equation gives 
the diffusive lengthscale along the magnetic field, where the diffusion 
proceeds rapidly. 

\begin{figure*}[ht]
\begin{center}
\centerline{\psfig{file=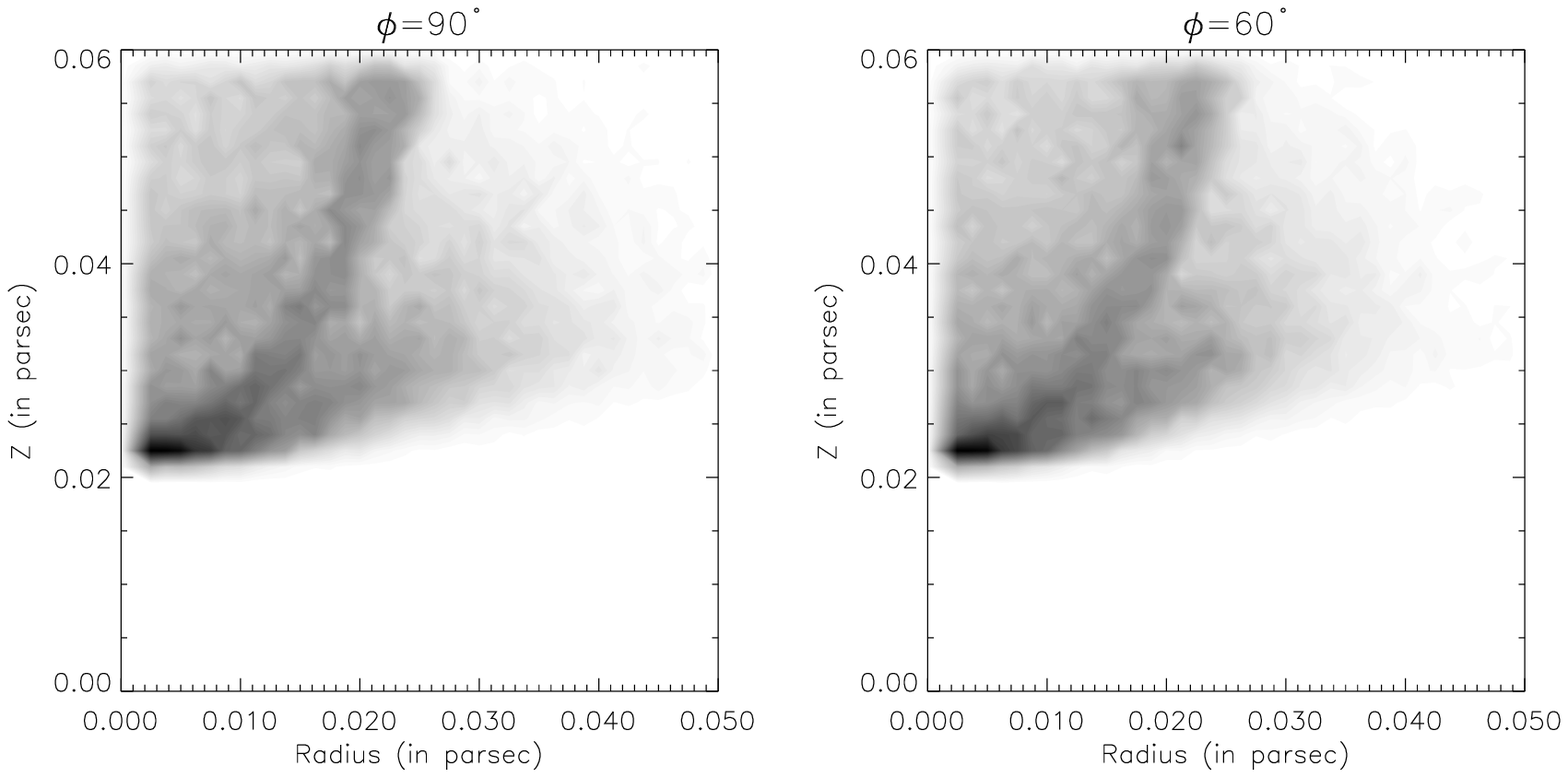,height=2.in}}
\centerline{\psfig{file=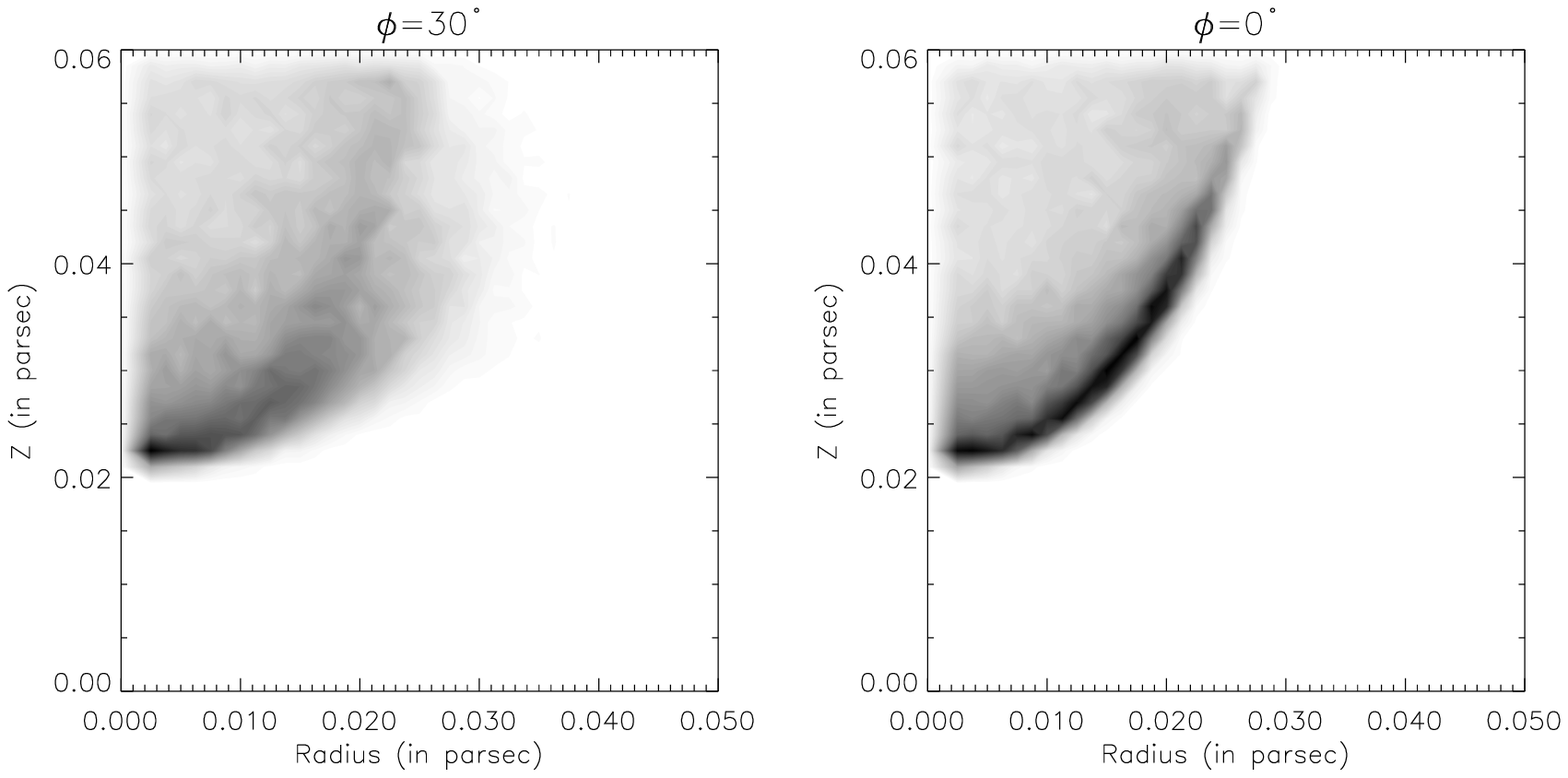,height=2.in}}
\end{center}
\caption{Synchrotron maps illustrating the effect of different angles of
observation with a gyro factor $\eta = 10$.}
\label{fig:2}
\end{figure*}

\subsection{Comparison with observations}
    
In this section we will use equation (\ref{Diffusion}) to determine lower
limits for
the gyro factor $\eta$ by matching $\Delta R_{\parallel}$ with the size of the 
rejuvenated shells observed in the SNRs CTB80 and G5.4-1.2. We make two 
assumptions which will lead to a {\it minimum} value 
for the gyro factor: 1) the timescale $t$ for the interaction between the pulsar
wind and the shell of the remnant is taken to be the age of the remnant; 2) the
diffusion process is assumed to take place in a uniform magnetic field rather
then in a curved, position dependent magnetic field. 

\subsubsection{CTB80 and PSR 1951+32}

We use as a reference Strom \& Stappers (2000), who observed this remnant at a
frequency of $\nu = 600$ MHz. Using a distance of 2 kpc, the rejuvenated shell 
has a size of $\sim$ 17 parsec. Taking the age of the system as 
$t = 10^5$ years, we obtain $\tau_{\rm loss}\simeq 8.2\times 10^7
B_{\mu{\rm G}}^{-3/2}$ years. Using equation (\ref{Diffusion}), we derive a
lower limit for the gyro factor in order to match this observation:
\begin{equation}
	\eta \ge 2.5\times 10^3 B_{\mu {\rm G}}^{3/2}.
\end{equation}

\subsubsection{G5.4-1.2 and PSR B1757-24}

We use Gaensler \& Frail (2000) as a reference for this remnant. Their map is
at a frequency of 330 MHz. The rejuvenated shell of the SNR has a size
of $\sim 30$ parsec for a distance of 4.0 kpc. By using a characteristic age
of $t=1.6\times 10^4$ year, we derive a value $\tau_{\rm loss}\simeq
1.1\times 10^8B_{\mu{\rm G}}^{-3/2}$ years. Using equation (\ref{Diffusion}) again, we
derive a lower limit for the gyro factor, which equals:
\begin{equation}
	\eta \ge 6.6\times 10^4 B_{\mu {\rm G}}^{3/2}.
\end{equation}

\section{Simulations}

In this section we show results from a Monte Carlo simulation, which traces the
propagation of test particles in the flow of a bow shock. We obtained the 
steady-state flow inside a bow shock by performing hydrodynamical simulations
with the Versatile Advection Code  
\footnote{See http://www.phys.uu.nl/\~{}toth/}(T\'oth 1996).

The test particles are contineously injected in the flow of this bow shock at
the site of the termination shock. We use It\^o stochastic differential
equations (SDEs), to simulate the random walk trajectories of these particles
in the flow of the bow shock (a detailed discussion can be found in the Ph.D. 
thesis of E. van der Swaluw \footnote{publicly available at: 
http://www.library.uu.nl/digiarchief/dip/diss/1967493/inhoud.htm}). 

By considering many realizations of the SDE in the bow shock flow, we obtain a
steady-state distribution of particles in phase space. We use this distribution
to produce a synchrotron map, which can illustrate the influence of a
gyro factor $\eta > 1$.

The hydrodynamical flow is axially symmetric around the Z-axis and the 
magnetic field strength is uniform {\it inside} the PWN. {\it 
Outside} the PWN, the magnetic field is uniform and directed perpendicular to
the pulsar velocity (see also figure 1).

Figure 2 shows the result for a case with $\eta =10$. In the upper left panel 
the line of sight of the observer is perpendicular with the magnetic field
lines outside the PWN. In the lower right panel the line of sight is parallel 
with the magnetic field lines, therefore the rejuvenated parts of the remnant
are {\it not} visible.

\section{Conclusions}

We have reconsidered the rejuvenation mechanism as proposed by Shull et al.
(1989). This was done by investigating the propagation of the injected
relativistic electrons at the site of the wind termination shock through the
PWN and the associated SNR shell. We conclude that a pulsar wind can rejuvenate
the shell of a SNR if the diffusive transport of the injected electrons is
strongly anisotropic, i.e. the gyro factor has to have a minimum value of
$\eta\sim 10^3-10^4$. The limits for the gyro factor derived in this work are
{\it lower} limits because of the assumptions we made about the interaction
time and the configuration of the magnetic field (section 2). Furthermore, our 
numerical experiment has shown that the line of sight between the observer and 
the magnetic field introduce a projection effect, which diminishes the
lengthscales of the rejuvenated parts of the SNR shell (section 3).

\acknowledgments

EvdS is currently supported by the European Commission under the TMR programme,
contract number ERB-FMRX-CT98-0168.

\end{document}